\documentclass{emulateapj}
\usepackage{amstext}
\usepackage{apjfonts}
\usepackage{amsmath}

\newcommand{\rhop}{\rho_{0 {\rm p}}}
\newcommand{\rhon}{\rho_{0 {\rm n}}}
\newcommand{\be}{\begin{equation}}
\newcommand{\ba}{\begin{array}}
\newcommand{\beqn}{\begin{eqnarray}}
\newcommand{\ee}{\end{equation}}
\newcommand{\ea}{\end{array}}
\newcommand{\eeqn}{\end{eqnarray}}
\newcommand{\bc}{\begin{center}}
\newcommand{\ec}{\end{center}}

\begin{document}
\submitted {Submitted to ApJL on June 2, 2003}
\journalinfo {Submitted to ApJL on June 2, 2003}

\shorttitle{NEUTRON-RICH MHD OUTFLOWS IN GRB SOURCES}
\shortauthors{VLAHAKIS, PENG, \& K\"ONIGL}

\title{Neutron-Rich Hydromagnetic Outflows in Gamma-Ray Burst Sources}

\author{Nektarios Vlahakis, Fang Peng, and Arieh K\"{o}nigl}
\affil{Dept. of Astronomy \& Astrophysics and Enrico Fermi
Inst., Univ. of Chicago, 5640 S. Ellis Ave., Chicago, IL 60637
\\ {vlahakis@jets.uchicago.edu, fpeng@oddjob.uchicago.edu, arieh@jets.uchicago.edu}
}

\begin{abstract}
We demonstrate that ``hot'' MHD outflows from neutron-rich black-hole debris
disks can significantly alleviate the baryon-loading
problem in gamma-ray burst (GRB) sources. We argue that the
neutron-to-proton ratio in disk-fed outflows might be as high
as $\sim 30$ and show, with the help of an exact semianalytic
relativistic-MHD solution, that the neutrons can decouple at a
Lorentz factor
$\gamma_{\rm d} \sim 15$ even as the protons continue to
accelerate to $\gamma_\infty \sim 200$ and end up acquiring $\sim 30\%$ of
the injected energy. We clarify the crucial role that the magnetic field
plays in this process and prove that purely hydrodynamic
outflows must have $\gamma_{\rm d} \gtrsim$ a few $\times 10^2$. The motion
of the decoupled neutrons is not collinear with that of the
decoupled protons, so, in contrast to previous suggestions based
on purely hydrodynamic models, the two particle groups
do not collide after decoupling. If the decoupled neutrons move
at an angle $\gtrsim 1/\gamma_{\rm d} \approx
3.8 \degr (15/\gamma_{\rm d})$ to the line of sight to the GRB
source, most of their emission after they decay into protons will remain unobservable.
\end{abstract}


\section{Introduction}
\label{introduction}
Gamma-ray burst (GRBs) are inferred to arise in relativistic
outflows of terminal Lorentz factor $\gamma_\infty \gtrsim 10^2$
\citep[e.g.,][]{LS01}. In the case of long-duration ($\Delta t
\gtrsim 2\ {\rm  s}$) bursts, there is growing evidence that the
outflows are highly collimated and involve a kinetic
energy $E_k \sim 10^{51}\ {\rm  ergs}$ 
\citep[e.g.,][]{PK02}. The outflows likely originate in
a newly formed neutron star (NS) or stellar-mass black hole (BH) and
are powered either by the NS or BH
spin or by the gravitational potential energy of a remnant accretion disk.
The jets can be driven either thermally or
magnetically. Magnetic fields offer a natural means of tapping
the rotational energy of the source and of guiding,
accelerating, and collimating the flow. In a recent series of
papers, Vlahakis \& K\"onigl [2001, 2003a (VK03a), 2003b (VK03b)] summarized the 
evidence in favor of magnetic driving as well as previous work on this topic, and
presented exact semianalytic solutions of the ``hot'' relativistic MHD
equations demonstrating that Poynting flux-dominated jets can transform
$\gtrsim 50\%$ of their magnetic energy into kinetic energy of
$\gamma_\infty \sim 10^2-10^3$ baryons.

The baryonic mass involved in long-duration GRB outflows is estimated to be
$M_{\rm p}=E_k/\gamma_\infty c^2 \approx 3 \times 10^{-6} (E_k/10^{51}\ {\rm  ergs})
(\gamma_\infty/200)^{-1}\, M_\sun$. This estimate poses an
apparent difficulty for realistic source models. For example, a
BH debris disk would need to be a factor $\sim 10^4$ more massive even
if 10\% of its gravitational potential energy could be converted
into outflow kinetic energy. A question thus arises as to how
the baryon loading of the jets remains so low even as
the energy deposition in the flow is highly efficient. A
promising proposal to alleviate this problem
was made by \citet*{FPA00}, who suggested that, if the source of
the outflow is neutron-rich, then the neutrons (which are only
very weakly affected by the electromagnetic field and are
accelerated primarily by collisional drag with ions) could
in principle decouple from the flow before the protons attain
their terminal Lorentz factor. If the mass source for the flow
is a NS then it will clearly be neutron-rich, and this
is evidently the case also for outflows fed by a BH
debris disk (see \S~\ref{disk}). However, it turns out that the
decoupling Lorentz factor $\gamma_{\rm d}$ in a thermally driven,
purely hydrodynamic (HD) outflow is of the order of the inferred
value of $\gamma_\infty$ (e.g., \citealt*{DKK99}; \citealt{B03b};
see \S~\ref{outflow}), which has so far limited the practical
implications of the \citet{FPA00} proposal.
In this Letter we demonstrate that a {\em hydromagnetic}
neutron-rich outflow can go a long way toward addressing the
baryon ``contamination'' problem. In \S~\ref{disk}
we consider the composition of GRB disks and construct an
illustrative disk model in the context of the supranova scenario
for GRBs. A semianalytic MHD solution representing an initially
neutron-rich outflow that undergoes decoupling at $\gamma_{\rm d} \ll
\gamma_\infty$ is presented in \S~\ref{outflow}. We summarize and discuss the
implications of our results in \S~\ref{implications}.

\section{Source Composition}
\label{disk}

\citet*{PWH03} reviewed the various models that attribute GRB
outflows to disks around newly formed stellar-mass BHs. They
considered disk accretion rates $\dot m_{\rm disk} \equiv [\dot
M_{\rm disk}/(M_\sun\ {\rm  s}^{-1})]=0.01-1$ and viscosity parameters
$\alpha=0.01-0.1$, and calculated typical midplane electron fractions
$Y_{\rm e}=\sum_j (Z_jX_j/A_j)$ (where $Z_j$, $X_j$, and $A_j$ are the
proton number, mass fraction, and atomic mass number, respectively) 
near the event horizon of a $3\, M_\sun$ BH in the range
$0.045-0.527$ (with $Y_{\rm e}$ decreasing with decreasing $\alpha$ for a given
$\dot m_{\rm disk}$ and increasing with decreasing $\dot m_{\rm disk}$ for a given
$\alpha$). It was argued that material reaching
the base of the outflow at the top of the disk would remain
neutron-rich if the midplane value of $Y_{\rm e}$ were
low. Similar results were presented by \citet{B03b}.

As a complement to this calculation, we now consider
debris disks of the type likely to arise in the
supranova scenario for GRBs \citep{VS98}. In this picture, a
supernova explosion leads to the formation of a rapidly
rotating, supramassive NS (SMNS), which, after losing rotational
support through a combination of electromagnetic and
gravitational radiation on a timescale of weeks to years, eventually 
collapses to a black hole (in the process triggering the
GRB). In the original proposal, the debris disk was envisioned
to represent the outer layers of an SMNS that
were left behind when its central regions collapsed. However, \citet{SBS00}
argued, on the basis of general-relativistic, 3D hydrodynamic
simulations of uniformly rotating stars with a
comparatively stiff equation of state (EOS), that the collapse
likely involves the entire SMNS. An alternative
possibility of forming a disk in stars with a stiff EOS is for
matter to be shed centrifugally in the equatorial plane as the
SMNS evolves (through the loss of angular momentum) toward the
gravitational-instability point. We illustrate this possibility by
considering the evolution of such an SMNS along the
mass-shedding limit from the maximum-mass point to the
maximum-angular-velocity ($\Omega$) point, where
it becomes unstable to collapse \citep*[see][]{CST94}. Using the EOS-M equilibrium
model results of \citet[]{CST94}, we infer that the
newly formed BH would have a (gravitational)
mass $M_{\rm BH}=2.04\, M_\sun$ and (assuming that none of the
shed mass became unbound) would be surrounded by a
debris disk of mass $M_{\rm disk}=0.06\, M_\sun$.\footnote{We neglect any
additional mass from the fallback of the original supernova
explosion that might surround the BH, as this material would not
be as highly magnetized as the SMNS and would therefore be
unlikely to partake in the GRB outflow.}

\citet{CST94} showed that, prior to the formation
of the BH, there always exists a region in the equatorial plane
outside the SMNS within which particles on circular orbits are
unstable to radial perturbations. We thus expect a bona fide
accretion disk to form only after the SMNS collapses, and we
approximate its initial extent as stretching from the innermost
stable circular orbit of the final BH ($\simeq 4GM_{\rm BH}
/c^2 \approx 12\ {\rm  km}$) to the equatorial radius ($R_{\rm e}\approx 22\ {\rm
km}$) of the Maximum-$M$ configuration
(i.e., $\Delta \varpi_{\rm i} \simeq 10\ {\rm  km}$, where $\varpi$ is
the cylindrical radius and the
subscript $i$ denotes the base of the flow). Most of the mass
would be initially concentrated near the radius of the maximum-$\Omega$ SMNS ($\simeq 16\
{\rm km}$, henceforth taken to be the typical value of $\varpi_{\rm i}$), so the disk
binding energy can be estimated as $E_{\rm b}
\approx GM_{\rm BH}M_{\rm disk}/\varpi_{\rm i} \approx 2 \times 10^{52}\ {\rm  ergs}$.
If we identify a characteristic burst duration of $10\ {\rm  s}$
with the accretion time of the
disk, we obtain a fiducial accretion rate of $\dot m_{\rm disk} =
0.006$. This rate is below the range considered by
\citet{PWH03} even as the characteristic midplane density
($\gtrsim 10^{13}\ {\rm g\ cm^{-3}}$) is much higher than  in
any of their models. Nevertheless, this disk and the inner
regions of most of the \citet{PWH03} models share the property that
their cooling is dominated by neutrino emission under marginally optically
thin conditions. If the outflows from these disks are strongly
magnetized, then much of the internal heating could be
due to Ohmic dissipation of wound-up magnetic field even
as the transport of angular momentum might be dominated by
large-scale magnetic stresses (see VK03b). If, however, we
represent both of these processes in terms of an
$\alpha$ viscosity, then a rough estimate of the parameter
$\alpha$ for the SMNS-spawned disk can be obtained by
equating the accretion time to the viscous angular-momentum
transport time in a neutrino-cooled disk model
\citep*[e.g.,][]{PWF99}, which yields $\alpha =
1.2\times 10^{-3}$. 

We can estimate the density of the innermost mass shell of the
maximum-$M$ SMNS that ends up in the disk by setting $M_{\rm
disk} = [4\pi R_{\rm e}^4 P_{\rm
s}(\rho_{\rm s})/{GM}][1-(2GM/R_{\rm e}c^2)]$ \citep{LRP93}, which
yields $P_{\rm s}= 1.5\times10^{32}\ {\rm dynes\ cm^{-2}}$. 
Based on the results of \citet{DH01}, the corresponding
density is $\rho_{\rm s} \approx 6\times 10^{13}\ {\rm g\
cm^{-3}}$, and the shell lies near the bottom of the inner crust
and is made up of free neutrons with $X_{\rm n} \simeq 0.8$ and of
neutron-rich composite nuclei with $Z \simeq 40$ and $Z/A \simeq 1/6$.
The disk mass thus comes mainly from the NS inner crust; the contribution
of the outer crust ($\sim 10^{-4}\, M_{\odot}$) is negligible.
After being shed from the NS, the composite nuclei
are dissociated and eventually end up releasing their
excess neutrons (with the remainder consisting mostly of
$^4$He). The exact composition of the disk-ejected material
is determined by nuclear reactions within the disk (but below the freezeout
point, where the nuclear reaction timescale comes to
exceed the vertical travel time of the ejected gas) as well as
by deuterium and helium production within the flow (e.g., \citealt{L02};
\citealt*{PGF02}). However, collision with neutrons just prior to
decoupling are expected to break apart
both $\alpha$ particles and deuterons (e.g., \citealt{PGF02};
\citealt{B03b}). In the limit that only free nucleons remain at
the time of decoupling, the value of $Y_{\rm e}$ for the NS-shed
material would be $\gtrsim 1/30$ and the neutron (n) to
proton (p) ratio would be $(1-Y_{\rm e})/Y_{\rm e} \lesssim 29$.
Although the n/p ratio at the base of the outflow would be lower
than this value, we expect it to be high enough to eventually
give rise to n-p drift speeds $\sim c$ that would lead to
decoupling and result in the collisional breakup of the bulk of the remaining
composite nuclei.\footnote{In the ``viscous heating and
optically thin neutrino cooling'' approximation to our
debris-disk model, the disk is vertically isothermal and has a
temperature $T_{\rm disk}=1.8\times 10^{10}\ {\rm K}$ and a midplane density
of $1.9\times 10^{13}\ {\rm g\ cm^{-3}}$ at $\varpi_{\rm i}$. For
the estimated value of $M_{\rm p}$ (see \S~\ref{introduction}) and with
$X_{\rm n}/X_{\rm p}=30$, the density at the bottom of the disk surface layer that
ends up in the flow is $\rho_{\rm disk}= 1.2\times 10^{11}\ {\rm g\ cm^{-3}}$ 
(corresponding to $\sim 0.1\%$ 
of $M_{\rm disk}$ being ejected). These values of $T_{\rm disk}$ and $\rho_{\rm
disk}$ are comparable to those used in the models of \citet{PWH03}.}
In the illustrative outflow solution presented in
\S~\ref{outflow} we use a constant pre-decoupling n/p ratio of 30.

\section{Flow Acceleration and Neutron Decoupling}\label{outflow}

The equations of motion for the neutron component of the outflow
are given by
\begin{eqnarray}\label{neutrons_momentum}
\left( \rhon U_{\rm n}^\kappa U_{\rm n}^\nu \right)_{, \nu} 
={\cal F}^\kappa \,,
\end{eqnarray}
where $\rhon $ is the rest-mass density of neutrons,
$U_{\rm n}^\kappa=  \left(\gamma_{\rm n} c \,, \gamma_{\rm n} {\boldsymbol V}_{\rm
n} \right)$ their four-velocity, and 
\be \label{collision_force}
{\cal F}^\kappa = \frac{\sigma_0 c }{2 m_{\rm p}} \rhop \rhon
\left(U^\kappa- U_{\rm n}^\kappa\right) 
\end{equation}
the neutron--proton collisional-drag force density.

The rest of the fluid -- consisting of protons (with rest-mass density $\rhop$),
radiation (with pressure $P_{\rm R} $), and e$^+$e$^-$ pairs
(with pressure $P_{\rm M} $) -- is moving with four-velocity
$U^\kappa= \left(\gamma c \,, \gamma {\boldsymbol
V}\right)$. The equations of motion of the charged fluid
component are
\begin{eqnarray}\label{protons_momentum}
\left[ \rhop U^\kappa U^\nu 
+ \left( P_{\rm R} + P_{\rm M}\right)\left(
4 \frac{ U^{\kappa} U^{\nu}}{c^2} +\eta^{\kappa \nu} \right)
\right]_{, \nu} = {\cal F}^\kappa_{\rm EM} -{\cal F}^\kappa ,
\end{eqnarray}
incorporating the p-n drag force density $-{\cal F}^\kappa$ and
the electromagnetic force density ${\cal F}^\kappa_{\rm EM}
= [ { {\boldsymbol J}\cdot {\boldsymbol E}}/{c}\,,
({J^0 {\boldsymbol{E}} +{\boldsymbol{J}} \times
{\boldsymbol{B}}})/{c} ]$. Here ${\boldsymbol E}\,,\,
{\boldsymbol{B}}$ is the electromagnetic field 
and $J^{\nu}=\left(J^0,\boldsymbol{J}\right)$ the associated
four-current (with $J^0/c$ representing the charge density).
As in VK03a, we assume that the photons possess
a blackbody distribution (under optically thick
conditions) and the pairs a Maxwellian distribution with zero
chemical potential. In the large-temperature limit ($\Theta \equiv
k_{\rm B}T/m_{\rm e}c^{2}\gg 1$)
the latter corresponds to a $4/3$ polytrope, $P_{\rm M} \propto
\Theta^4 \propto P_{\rm R}$; in this case $P_{\rm M}/P_{\rm R} =
180/\pi^4 \approx 1.85$ is a constant, so the matter and
radiation can be treated as a single fluid.

The system (\ref{neutrons_momentum})--(\ref{protons_momentum}), 
together with Maxwell's laws and the proton and neutron mass
conservation relations can in principle be solved to yield the velocities of
the charged ($U^\kappa$) and neutral ($U_{\rm n}^\kappa$) components 
as well as the electromagnetic field (${\boldsymbol E}\,,\, {\boldsymbol{B}}$)
and the thermodynamic quantities ($\rhop\,, \rhon\,,
\Theta$). However, a simpler method that is sufficient for our
purposes is described below.

During the initial (pre-decoupling) outflow phase, the
neutrons and protons are well coupled and their velocities are
comparable: $U_{\rm n}^\kappa \approx U^\kappa$.
By eliminating the drag force between equations
(\ref{neutrons_momentum}) and (\ref{protons_momentum}) and
setting $U_{\rm n}^\kappa \approx U^\kappa$, we end up
with the same system of equations as in the pure proton/lepton
case ---  we
only need to replace $\rho_0$ by $\rhon+\rhop=\rhop(1+\rhon/\rhop)$.
By applying the same $r$ self-similar model as in VK03a, 
we can study the neutron-rich outflow and evaluate all the
physical quantities, including $U^\kappa$. Next, equation 
(\ref{neutrons_momentum}) with $U_{\rm n}^\kappa \approx U^\kappa$
(or eq. [\ref{protons_momentum}]) yields the force ${\cal F}^\kappa$. 
Finally, equation (\ref{collision_force}) gives the drift
velocity, which can be written (with the help of the continuity
equation) as
\be
\label{driftv}
U_{\rm n}^\kappa - U^\kappa \approx
-2 m_{\rm p} U^\nu U^\kappa_{, \nu}/\sigma_0 c \rhop\,.
\end{equation}
So long as the drift velocity is much smaller than the proton
velocity, the neutrons remain well coupled to the protons.
In the general (nonradial) case there are two possible modes of
decoupling, which can be dubbed ``along the poloidal field'' and ``in the transfield
direction.'' The former happens when the relative velocity between
protons and neutrons along the poloidal magnetic field,
$V_\parallel - V_{{\rm n} \parallel}$, grows to $\sim
V_\parallel \approx c$, and the latter when $V_{{\rm n} \bot}$
becomes $\sim \Delta  \varpi \cos \vartheta / \gamma \tau_{\rm dyn}$, where
$\vartheta $ is the jet opening half-angle,
$\Delta \varpi \cos \vartheta $ is its width,
and $\tau_{\rm dyn} \sim z/\gamma c$ is the dynamical timescale.
(The subscripts $\parallel\,,\, \bot$ denote components of a
vector along the poloidal
magnetic field and in the transfield direction, respectively.)
\begin{center}
\plotone{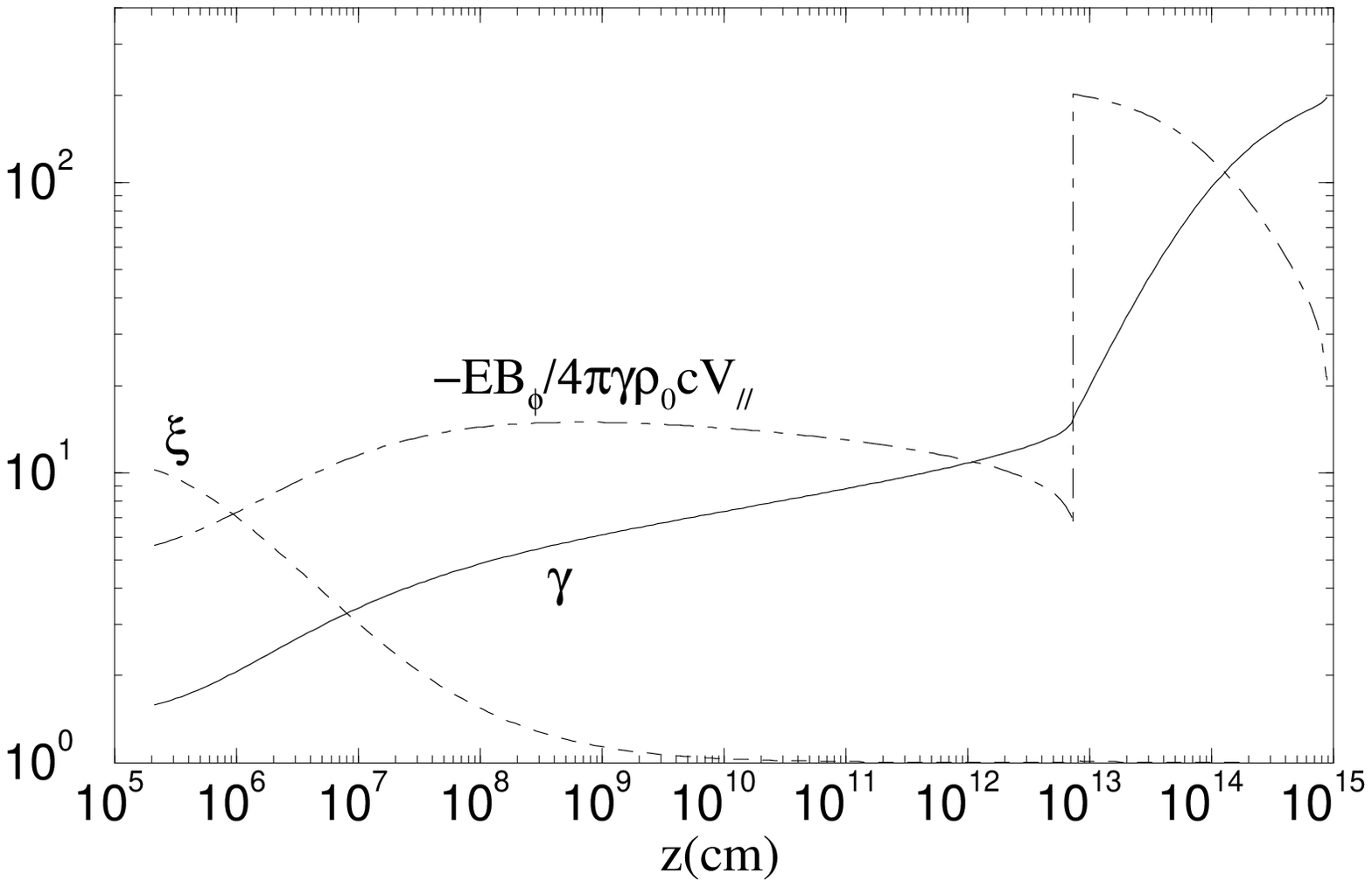}
\figcaption[]
{Lorentz factor $\gamma$ ({\emph{solid}}), specific enthalpy 
$\xi = (\rho_0 c^2 + 4P_{\rm R}+4 P_{\rm M})/\rho_0 c^2$
({\emph{dashed}}), and Poynting-to-mass flux ratio
$-(c/4 \pi ) E B_\phi / \gamma \rho_0 c^2 V_\parallel$
({\emph{dot-dashed}}) as functions of height ($z$) along the innermost fieldline.
Here $\rho_0 =\rhon+\rhop$ below the decoupling point and
$\rho_0=\rhop$ above it. The baryon mass-normalized quantities
are discontinuous at the decoupling transition, where they
increase by a factor $1+\rhon/\rhop$.
\label{fig1}}
\end{center}

We model the post-decoupling evolution of the charged fluid component
using the same $r$ self-similar model used in the
pre-decoupling phase (but replacing $\rho_0$ by $\rhop$).
Figures \ref{fig1} and \ref{fig2} show a solution representing
an outflow from a disk with an inner radius
$r_{\rm in}=1.2\times 10^6\ {\rm cm}$ and outer radius $r_{\rm out} = 2 r_{\rm
in}$. The initial conditions for the
innermost fieldline ($r_{\rm i}=r_{\rm in}$, $\theta_{\rm i}=80 \degr$)
are: $\rhon/\rhop=30$, $\rhon+\rhop=10^5\ {\rm g\ cm^{-3}}$,
$\Theta=2.2$, $B_{p}=10^{12}\ {\rm G}$, $B_\phi=-10^{14}\ {\rm G}$,
$\vartheta=55 \degr$, $V_p=0.6916 c$, and $V_\phi=0.35
c$.\footnote{The value of the model parameter $F$ is
1.05 before decoupling and 0.1 after decoupling.
We use the form of the n-p collisional cross section given in
\citet{DKK99}, with $\sigma_0=3 \times 10^{-26}\ {\rm cm}^2$.}
These conditions correspond to a rough equipartition between
comoving magnetic and thermal pressures --- 
$(B^2-E^2) / 8 \pi (P_{\rm R}+P_{\rm M})=1.0$,
a total outflowing baryon mass $M_{\rm b}=2 \int \! \! \! \! \int
\gamma \rho_0 {\boldsymbol{V}} \cdot d {
\boldsymbol {S}} \ \Delta t= 9.22 \times 10^{-5} (\Delta
t/10 \mbox{s}) M_\sun$, and total energy ${\cal E}_{\rm i} = 3.62
\times 10^{51} (\Delta t / 10\ \mbox{s})\ {\rm ergs}$.
The various components of the initial total energy are
$[$electromagnetic, enthalpy (including rest energy),
kinetic$]=[5.615,\ 15.631,\ 0.583] \times  M_{\rm b} c^2$.
During the pre-decoupling phase,
part of the enthalpy is converted into baryon kinetic energy
(resulting in an increase in $\gamma$) and another is converted
into electromagnetic energy (resulting in an increase in the
Poynting-to-mass flux ratio; see Fig.~\ref{fig1}). Above
$z\approx 10^9\ {\rm cm}$ the fluid is practically cold ($P_{\rm
R}+P_{\rm M} \ll \rho_0 c^2 \Leftrightarrow \xi \approx 1$) and the magnetic field is
starting to accelerate the matter: in this regime the increasing
$\gamma$ corresponds to a decreasing Poynting-to-matter flux ratio.
\begin{center}
\plotone{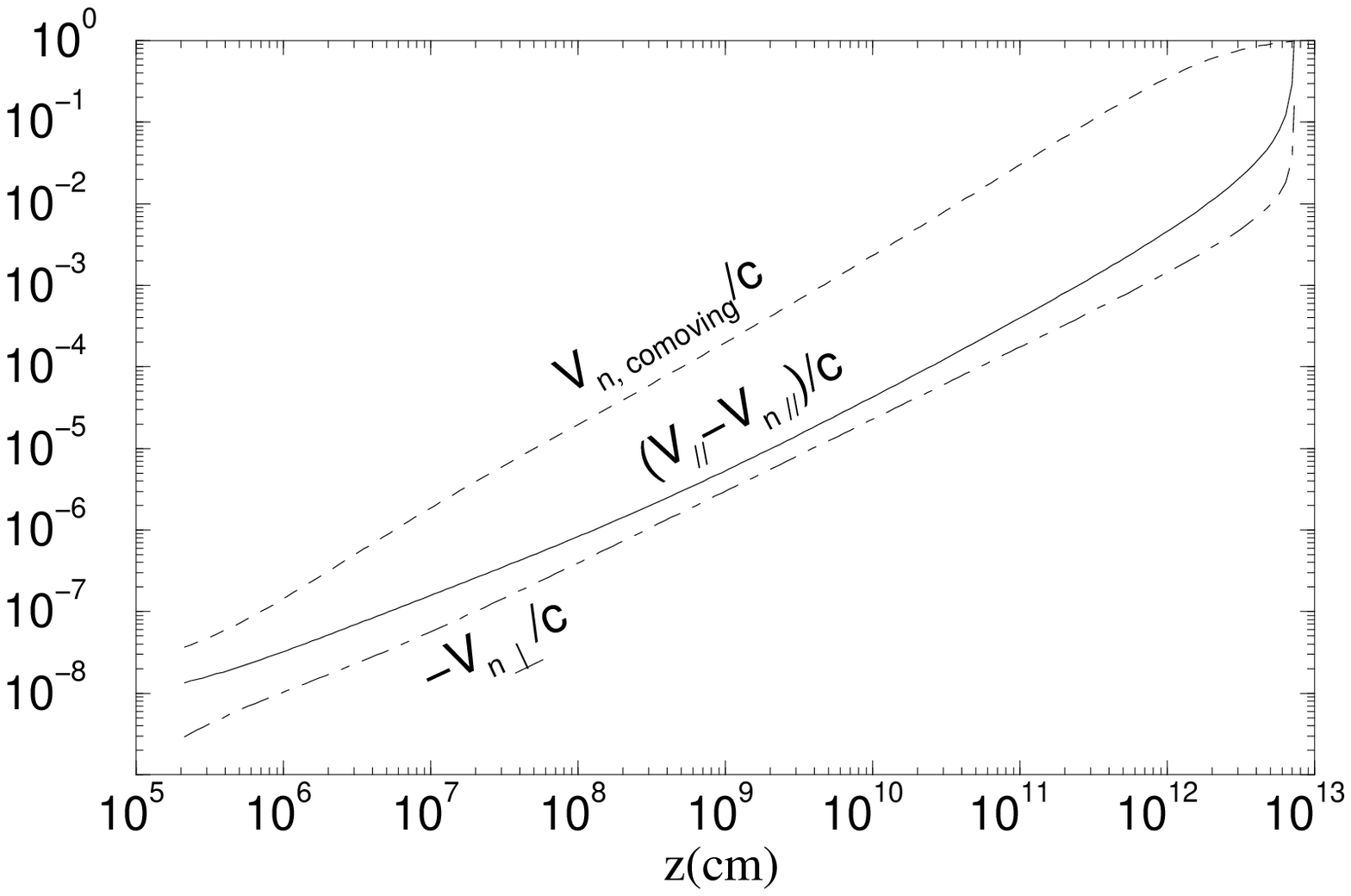}
\figcaption[]
{Proton--neutron drift velocity along the poloidal magnetic field
({\emph{solid}}),
in the transfield direction ({\emph{dot-dashed}}),
and in the comoving proton frame ({\emph{dashed}}) for the
outflow described in Fig.~\ref{fig1}.
\label{fig2}}
\end{center}

Figure \ref{fig2} shows the various components of the drift
velocity. At $z = z_{\rm d} \approx 7 \times 10^{12}\ {\rm cm}$, $V_\parallel - V_{{\rm n}
\parallel} \sim c$ and the neutrons decouple from the protons. 
In this particular solution the ``along the flow'' decoupling
happens first, although at the time of decoupling the neutrons
also have a nonnegligible outward-directed transfield velocity
component ($V_{{\rm n} \bot} \sim 0.1 c \ll V_{{\rm
n} \parallel}$).
The latter component arises from the ongoing magnetic
collimation of the jet: by the time the flow reaches $z_{\rm d}$,
$\vartheta$ is already decreased to $\sim 7
\degr$.\footnote{The value of $\vartheta$ for the proton
component changes little after decoupling; to obtain the exact value of the angle
between the two particle species at the time of decoupling, the system of equations
(\ref{neutrons_momentum})--(\ref{protons_momentum}) must be
solved without the approximation $U_{\rm n}^\kappa \approx U^\kappa$.}  
At the decoupling point $\gamma=\gamma_{\rm d} \approx 15.3$, and
the energy of the charged fluid component is
${\cal E}_{\rm c,d} = 1.16 \times 10^{51} (\Delta t / 10\ \mbox{s})\
{\rm ergs} = 32 \% \ {\cal E}_{\rm i}$
[with the remaining $68 \%$ residing in the neutron kinetic and rest energies,
$\gamma_{\rm d} M_{\rm b} c^2 / (1+\rhop/\rhon)$].
In units of the proton rest energy $M_{\rm p} c^2 = M_{\rm b} c^2 / (1+\rhon/\rhop)$, 
the various components of this energy are
$[$electromagnetic, enthalpy (including rest energy), kinetic$]=[201.79,\ 1.22,\ 14.31]
\times M_{\rm p} c^2 $. 

Above $z_{\rm d}$ we continue the integration 
of equation (\ref{protons_momentum}) with ${\cal F}^\kappa=0$.
Figure \ref{fig1} shows that the acceleration in this regime
is magnetic: the electromagnetic energy decreases and 
the Lorentz force accelerates the protons to a terminal Lorentz factor
$\gamma_\infty = 200$.
The final proton kinetic energy is
$E_k = \gamma_\infty M_{\rm p} c^2 \approx {\cal E}_{\rm c,d}$,
indicating that the acceleration efficiency in
the post-decoupling region is $\approx 100\%$. 

The above solution demonstrates that, in contrast to the HD case
considered by previous authors, neutron decoupling in an MHD
outflow can occur with $\gamma_{\rm d} \ll \gamma_\infty$. To
understand why magnetic acceleration is crucial to this outcome,
consider approximating the local scaling of the Lorentz factor
with cylindrical radius by $\gamma \propto \varpi^{\delta_1}$ and
the fieldline (or streamline) shape by $r \approx z \propto
\varpi^{1+\delta_2}$, where $\delta_1\,,\, \delta_2$ are positive constants.
Under the approximation of exact mass conservation for the
proton component, $\gamma \rhop \propto
\varpi^{-2}$. At the decoupling point $\tau_{\rm coll} \approx \tau_{\rm
dyn}$, where $\tau_{\rm coll} = m_{\rm p} / \sigma_0 c
\rhop$.\footnote{A more exact condition for the decoupling comes from
eq. (\ref{driftv}), which implies $V_\parallel - V_{{\rm n}\parallel} 
\sim c \Leftrightarrow 2 (m_{\rm p} / \sigma_0 \rhop) d\gamma/dz \sim 1 $, or
$\tau_{\rm dyn} \sim (2 d \ln \gamma / d \ln z)\, \tau_{\rm coll}$.}
Using the above scalings, we obtain
$\gamma_{\rm d} = \gamma_{\rm i} 
(\rho_{0 {\rm p, i}} r_{\rm i} \sigma_0 / m_{\rm p}
\gamma_{\rm i})^{\delta_1/(2 \delta_1 +1 - \delta_2)}$ and
$r_{\rm d}=r_{\rm i} (\gamma_{\rm d} / \gamma_{\rm i})^{(1+\delta_2)/\delta_1 }$.
In the HD case, the energy equation $\xi \gamma \propto \Theta
\gamma =const$ and the polytropic relation $\Theta^4 \propto \rhop^{4/3}$
always imply a linear increase of $\gamma$ with
$\varpi$. Thus, $\delta_1 =1$ in this case, and the minimum value of
$\gamma_{\rm d}$,  $\gamma_{\rm d\, HD\, min} = \gamma_{\rm i}
(\rho_{0 {\rm p, i}} r_{\rm i} \sigma_0 / m_{\rm p} \gamma_{\rm i})^{ 1/3}$,
is attained in a conical flow ($\delta_2=0)$.
For typical values of the source size (at least $10^6\ {\rm cm}$)
and initial proton density\footnote{As a rough estimate, $\rho_{0 {\rm p, i}}\sim
M_{\rm p} / 4 \pi r_{\rm i}^2 c \Delta t$, 
with the ejected proton mass inferred from $M_{\rm p}=E_k /\gamma_\infty
c^2$ using $E_k \sim 10^{51}\ {\rm ergs}$ and
$\gamma_\infty =$ a few $\times 10^2$.} (at
least a few times $10^2\ {\rm g\ cm^{-3}}$),
$\gamma_{\rm d\, HD\, min}$ is a few times $10^2$.
On the other hand, in MHD flows where a significant part of the
enthalpy can be initially transferred into Poynting energy,
the pre-decoupling acceleration is slower, corresponding to
$\delta_1 <1$. (In the solution presented here, $\delta_1
\approx 0.13$ and $\delta_2\approx 0.15$.)
This makes it possible for  $\gamma_{\rm d}$
to be $\ll \gamma_{\rm d\, HD\, min}$. 
The energy deposited in the Poynting flux is subsequently
returned to the matter as kinetic energy, enhancing the fraction
of the total energy used to accelerate the proton component.

\section{Summary and Implications}
\label{implications}
We argued that GRB outflows emanating from accreting debris
disks around stellar-mass black holes could possess a high ($\lesssim 30$)
pre-decoupling neutron-to-proton mass ratio, and we demonstrated
semianalytically, by means of a ``hot'' relativistic-MHD solution, that n-p
decoupling can occur at a Lorentz factor ($\gamma_{\rm d} \approx 15$ in the
presented example) that is significantly lower than the terminal
$\gamma$ of the protons ($\gamma_\infty = 200$ in
our solution). We explained why, in contrast, $\gamma_{\rm d}$ cannot be smaller
than a few times $10^2$ if the acceleration is purely
hydrodynamic. In the outflow solution that we constructed, the
protons, while constituting only 
$\sim 3\%$ of the ejected mass, end up
with a kinetic energy $\sim 10^{51}\ {\rm ergs}$ that represents
$\sim 1/3$ of the injected energy (with the remainder going to neutron kinetic energy).

This could significantly alleviate the baryon-loading problem
in GRB source models. For example, in the illustrative
SMNS-shed disk model that we considered, the outflow rate represents
$\sim 0.1\%$ of the mass inflow rate. If a fraction of this order of
the disk mass can be ejected (with $X_{\rm n}/X_{\rm
p}$ that becomes $\sim 30$ just before decoupling) in a ``hot'' magnetized jet,
then it may not be necessary to consider alternative sources of
energy or mass for GRB outflows, for which the converse problem (how
to avoid having too few baryons) is often encountered \citep[e.g.,][]{LE03}.

The decoupled neutrons decay into protons at a
distance $r_\beta=\gamma_{\rm d} c \tau_\beta \approx 4\times 10^{14}
(\gamma_{\rm d}/15)\ {\rm cm}$, where $\tau_\beta \approx 900\ {\rm
s}$ is the comoving decay time. Nonthermal shock-induced emission from the resulting
charged-particle shell may be hampered by the lack
of a strong magnetic field unless significant field amplification
occurs in the shocks themselves \citep[e.g.,][]{B03a} or if the
neutrons propagate within a highly magnetized medium (as might
happen in the supranova scenario; \citealt{KG02}). 
Internal collisions arising from the same velocity perturbations
that are invoked to account for the observed $\gamma$-ray
emission in the internal-shock scenario would typically occur on a scale $\ll
r_\beta$ and thus (in contrast to the situation in the proton
shell) would not give rise to radiative shocks.
Emission might, however, be expected from the reverse shock driven
into the decaying neutron shell as it starts to decelerate. 
Most of this shell's kinetic energy would, however, be radiated from the
external shock that is driven into the ambient gas. The bulk of the radiation would
be emitted at the distance $r_{\rm dec,n}$ where the mass swept
up from the ambient medium comes to exceed $\sim 1/\gamma_{\rm
d}$ of the shell mass.
This would occur on a timescale $\sim r_{\rm dec,n}/2\gamma_{\rm d}^2 c$, 
which, for a uniform environment, is a factor 
$\sim 10^3\, (X_{\rm n}/30 X_{\rm p})^{1/3}(\gamma_\infty/13\gamma_{\rm d})^{7/3}$
larger than the corresponding timescale for the original proton shell.
Most of this emission would, however, remain undetectable
if the angle between the neutron-shell velocity and the line of
sight exceeded $\sim 1/\gamma_{\rm d} \approx 3.8 \degr
(15/\gamma_{\rm d})$. It is in any case
clear that the transverse n-p relative motion at the time of
decoupling precludes any subsequent interactions between the two
shells (see, e.g., \citealt{PD02} and \citealt{B03a}). A detailed
consideration of the possible observational signatures of the
neutron shell remains an interesting problem for future research.

\acknowledgements We thank Andrei Beloborodov, Gregory Cook, Jason Pruet, and Stuart
Shapiro for helpful discussions and correspondence. This work
was supported in part by NASA grant NAG5-12635.

\end{document}